\documentclass[preprint,showpacs,preprintnumbers,amsmath,amssymb]{revtex4}
\usepackage{graphicx, epsfig}
\begin{document}

\title{Correlation between Local Structure Distortions and Martensitic Transformation in Ni-Mn-In alloys}
\author{D. N. Lobo$^1$, K. R. Priolkar$^1$, P. A. Bhobe$^{2,3}$, D. Krishnamurthy$^4$ and S. Emura$^4$}
\address{$^1$Department of Physics, Goa University, Taleigao-Plateau, Goa-403 206 India}
\address{$^2$Institute for Solid State Physics, The University of Tokyo, Kashiwa, Chiba 277-8581, Japan}
\address{$^3$RIKEN, SPring-8 Centre, Sayo-cho, Sayo-gun, Hyogo 679-5148, Japan}
\address{$^4$Institute of Scientific and Industrial Research, Osaka University, 8-1 Mihogaoka, Ibaraki,Osaka 567-0047, Japan}

\date{\today}

\begin{abstract}
The local structural distortions arising as a consequence of increasing Mn content in
Ni$_{2}$Mn$_{1+x}$In$_{1-x}$ (x=0, 0.3, 0.4, 0.5 and 0.6) and its effect on martensitic transformation have been
studied using Extended X-ray Absorption Fine Structure (EXAFS) spectroscopy. Using the room temperature EXAFS at
the Ni and Mn K-edges in the above compositions, the changes associated with respect to the local structure of
these absorbing atoms are compared. It is seen that in the alloys exhibiting martensitic transformation ($x \ge
0.4$) there is  a significant difference between the Ni-In and Ni-Mn bond lengths even in the austenitic phase
indicating atomic volume to be the main factor in inducing martensitic transformation in Ni-Mn-In Heusler alloys.
\end{abstract}

\maketitle

Non stoichiometric Ni-Mn-Z (Z = Sn and In) alloys exhibiting martensitic transformation in ferromagnetic state
have been studied for their many novel properties. These alloys exhibit giant reverse magneto-caloric effect
\cite{krenke-natmat,preeti,ali,du,liu,khan,dub}, large magnetic field induced strains
\cite{kainuma,kainuma-nat,kainuma2}, magnetic superelasticity \cite{krenke} due to improved possibility of
driving structural transformation by magnetic fields.

The martensitic transformation in these alloys has been a subject of debate in literature. Stoichiometric
Ni$_2$MnIn has a stable L2$_1$ crystal structure. Increasing substitution of In by Mn leads to martensitic
instability with the transformation temperature $T_M$ steadily increasing with Mn concentration \cite{acet}. A
simplistic explanation for such a behavior is the increase in electron per atom (e/a) ratio. However, this is not
entirely convincing as Ni-Mn-In alloys have higher $T_M$ than Ni-Mn-Sn alloys for a given Mn content inspite of
the former having lower e/a ratio. Another interpretation that has emerged, after systematic studies on
Ni$_2$Mn$_{1+x}$Z$_{1-x}$ with Z = Ga, In, Sn and Sb is based on the atomic volume or the chemical pressure
effect. According to this, the martensitic transformation occurs in alloys having austenitic lattice parameter,
$a_{cubic} \approx 6.0$\AA~ or less \cite{planes}. This is important because the magneto-thermal and
magneto-mechanical properties of these alloys have the same physical origin. The magnetic shape memory and
magnetocaloric properties of these alloys are strongly connected to the martensitic transition.  There exists an
intimate relationship between structural and magnetic degrees of freedom in these alloys. Therefore there is a
need to understand details of martensitic transformation to gain an insight of the magneto structural
relationship and thereby into microscopic principles governing magnetic shape memory effect and magnetocaloric
effect.

Recent EXAFS studies on Ni$_{50}$Mn$_{35}$Sn$_{15}$ \cite{bhobe1} and Ni$_{50}$Mn$_{35}$In$_{15}$ \cite{bhobe2}
reveal a possibility of existence of local structural distortions even in the austenitic phase. However, in these
studies there was no comparison made with stoichiometric alloys or compositions that do not undergo martensitic
transformation. Therefore it is still not clear if the effect of Mn substitution leads only to a overall
contraction of the lattice or there are some local structural distortions which play a more important role in
inducing martensitic transformation.

In this letter, we focus our attention on the local structure around Ni and Mn in Ni$_{2}$Mn$_{1+x}$In$_{1-x}$
($0 \le x \le 0.6$) to gain a better understanding of microscopic effects responsible for martensitic
transformation these magnetic shape memory alloys.

Polycrystalline Ni$_2$Mn$_{1+x}$In$_{1-x}$ with $x$ = 0, 0.3, 0.4, 0.5 and 0.6 were prepared by arc-melting the
starting elements ($\ge$ 99.99\% purity) under argon atmosphere. To ensure good homogeneity, the ingots were
flipped over and re-melted 4-5 times with a total weight loss of $\le$ 1\% and annealed at the temperature of
1000 K for 48 h in an evacuated quartz ampoule followed by quenching in ice cold water. The samples were
characterized X-ray diffraction, Energy dispersive analysis by X-rays (EDX), differential scanning calorimetry
(DSC) and magnetization for their structure, composition and magnetic and martensitic properties. EXAFS at Ni K
and Mn K edges were recorded at Photon Factory using beamline 12C at room temperature. For EXAFS measurements the
samples to be used as absorbers, were ground to a fine powder and uniformly distributed on a scotch tape. These
sample coated strips were adjusted in number such that the absorption edge jump gave $\Delta\mu t \le 1$ where
$\Delta\mu$ is the change in absorption coefficient at the absorption edge and $t$ is the thickness of the
absorber. The incident and transmitted photon energies were simultaneously recorded using gas-ionization chambers
as detectors. Measurements were carried out from 300 eV below the edge energy to 1000 eV above it with a 5 eV
step in the pre-edge region and 2.5 eV step in the EXAFS region. At each edge, at least three scans were
collected to average statistical noise. Data analysis was carried out using IFEFFIT \cite{Newville} in ATHENA and
ARTEMIS programs \cite{Ravel}. Here theoretical fitting standards were computed with FEFF6
\cite{Ravel2,Zabinsky}. The data in the k range of (2 - 12) \AA$^{-1}$ and R range 1 to 3 \AA~ was used for
analysis.

Ferromagnetic ordering temperature, T$_c$ and T$_M$ obtained from magnetization and DSC measurements for the
alloys along with their compositions obtained from EDX are listed in Table \ref{character}. These values agree
well with those reported in literature \cite{acet}.
\begin{table}
\caption{\label{character} Compositions, T$_c$ and T$_M$ of Ni$_2$Mn$_{1+x}$In$_{1-x}$ alloys determined from
EDX, magnetization and DSC measurements}
\begin{tabular}{lccccc}
\hline

Sample & Ni & Mn & In & T$_c$ (K) & T$_M$ (K) \\
Ni$_2$MnIn & 50.6 & 26.3 & 23.1 & 316 &--\\
Ni$_2$Mn$_{1.3}$In$_{0.7}$ & 50.2 & 32.7 & 17.1 & 315 & -- \\
Ni$_2$Mn$_{1.4}$In$_{0.6}$ & 50.8 & 34.2 & 15.0 & 314 & 270\\
Ni$_2$Mn$_{1.5}$In$_{0.5}$ & 51.1 & 36.7 & 12.2 &  -- & 390\\
Ni$_2$Mn$_{1.6}$In$_{0.4}$ & 49.9 & 40.2 & 9.9 & -- & 500\\
\hline
\end{tabular}
\end{table}

\begin{figure}[h]
\includegraphics[scale=0.75]{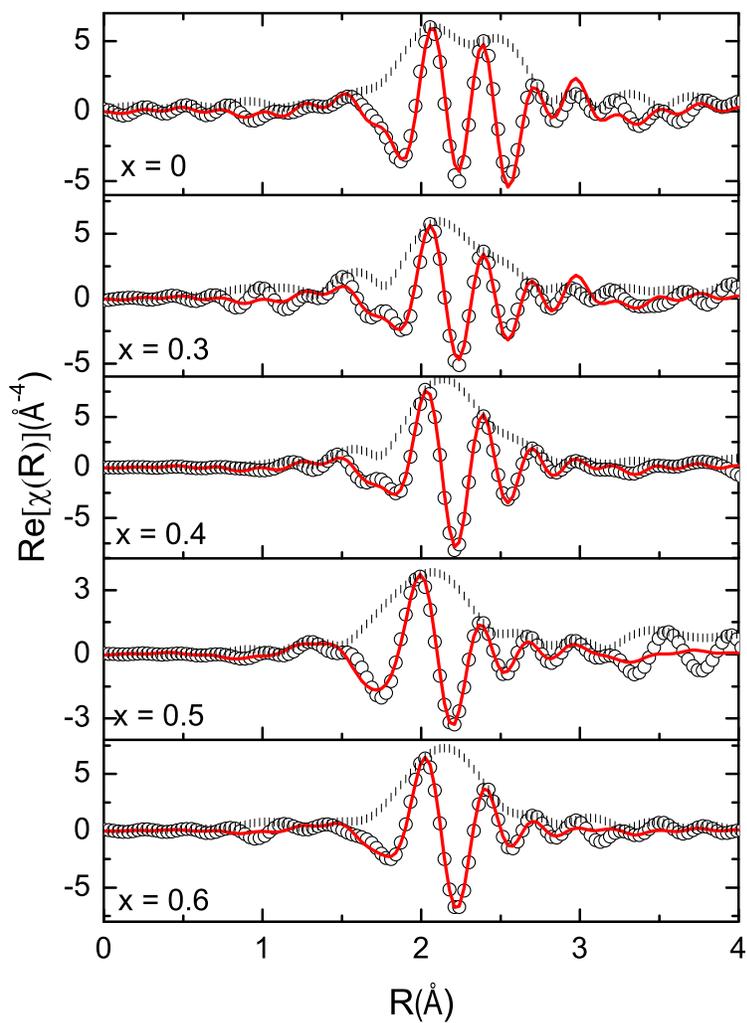}
\caption{\label{FT_mag all}  Real part (circles) and magnitude (vertical lines) of Fourier transform  of EXAFS
spectra at the Ni K-edge in Ni$_2$Mn$_{1+x}$In$_{1-x}$. The solid line is the best fit to the data.}
\end{figure}

The  magnitude and real component of Fourier transform (FT) of the EXAFS spectra at the Ni K-edge for all
Ni$_2$Mn$_{1+x}$In$_{1-x}$ along with fitted curves are shown in Fig. \ref{FT_mag all}. The spectra shown are
uncorrected for phase difference, however the bond length values reported herein are phase corrected values. A
dual peak structure is visible in the magnitude of FT in case of Ni$_2$MnIn ($x$ = 0). This dual peak structure
is due to backscattering from Mn and In atoms which surround the Ni atom. Since the phase difference introduced
by the two atoms is different a dual peak structure results inspite of the two (Mn and In) being equidistant from
the central absorbing atom (Ni). With increasing Mn substitution in place of In ($x > 0$), the dual peak
structure merges into a single broad peak.

In the austenitic phase, Ni occupies the body centered position in the B2 sub-cell of the L2$_1$ unit cell.
Therefore Ni should be equidistant from Mn (Y-site) and In/Mn (Z-site). Indeed in case of Ni$_2$MnIn, Ni-Mn and
Ni-In bond distances obtained from fitting Ni EXAFS are exactly equal. For all other compositions, differences
between  Ni-Mn and Ni-In bond lengths are obtained with Ni-Mn bond distance being always shorter than Ni-In bond
distance. This is very clear from the bond length values reported in Table \ref{bond}. The alloys with $x$ = 0
and 0.3 do not undergo martensitic transformation while the $x$ = 0.4 alloy undergoes martensitic transformation
below room temperature. Yet there is a difference between the Ni-Mn and Ni-In bond distances. Therefore, the
distortions in the non-stoichiometric alloys are of local nature which are present even in the austenitic phase
where the structure is ordered L2$_1$. Furthermore, the bond distances obtained from Mn K-edge EXAFS also support
this point. The Mn-Ni bond distances are in agreement with Ni-Mn bond lengths obtained from Ni EXAFS. Secondly,
the Mn-In/Mn bond distances are smaller than Ni-Ni bond distances obtained from Ni EXAFS in all compositions
except x = 0 indicating the presence of local distortion around the Mn site. In L2$_1$ ordered structure, Ni-Ni
or Mn-In bond length should be half the unit cell length and hence should be equal.  This is true in case of
Ni$_2$MnIn but in case of alloys having excess Mn composition the Mn-In/Mn bond lengths are shorter. Considering
the fact the alloys have L2$_1$ structure, the above differences in bond distance would imply that the Mn
occupying In site is displaced from its normal crystallographic site causing a local structural disorder.
\begin{table}
\caption{\label{bond}Bond distances obtained from analysis of Ni and Mn K edge EXAFS in
Ni$_2$Mn$_{1+x}$In$_{1-x}$ alloys. The numbers in parenthesis indicate uncertainty in the last digit.}
\begin{tabular}{lcccccc}
\hline Sample & \multicolumn{3}{c}{Ni EXAFS} & \multicolumn{3}{c}{Mn EXAFS}\\
&\multicolumn{6}{c}{Bondlength (\AA)}\\
&Ni-Mn & Ni-In & Ni-Ni & Mn-Ni & Mn-In & Mn-Mn \\
\hline
Ni$_2$MnIn &   2.626(3) &  2.626(3) &  3.04(1) & 2.618(5) & 3.02(1) & --\\
Ni$_2$Mn$_{1.3}$In$_{0.7}$ &   2.596(4) &  2.620(3) &  3.03(1) & 2.592(4) & 2.94(2) & 2.91(2)\\
Ni$_2$Mn$_{1.4}$In$_{0.6}$ &   2.565(3) &  2.620(3) &  3.02(1) & 2.562(3) & 2.89(1) & 2.93(1)\\
Ni$_2$Mn$_{1.5}$In$_{0.5}$ &   2.562(5) &  2.602(4) &  3.01(1) & 2.565(5) & 2.88(1) & 2.93(2)\\
Ni$_2$Mn$_{1.6}$In$_{0.4}$ &   2.557(4) &  2.600(3) &  3.00(1) & 2.574(4) & 3.03(1) & 2.95(1)\\
\hline
\end{tabular}
\end{table}

\begin{figure}[h]
\includegraphics[scale=1]{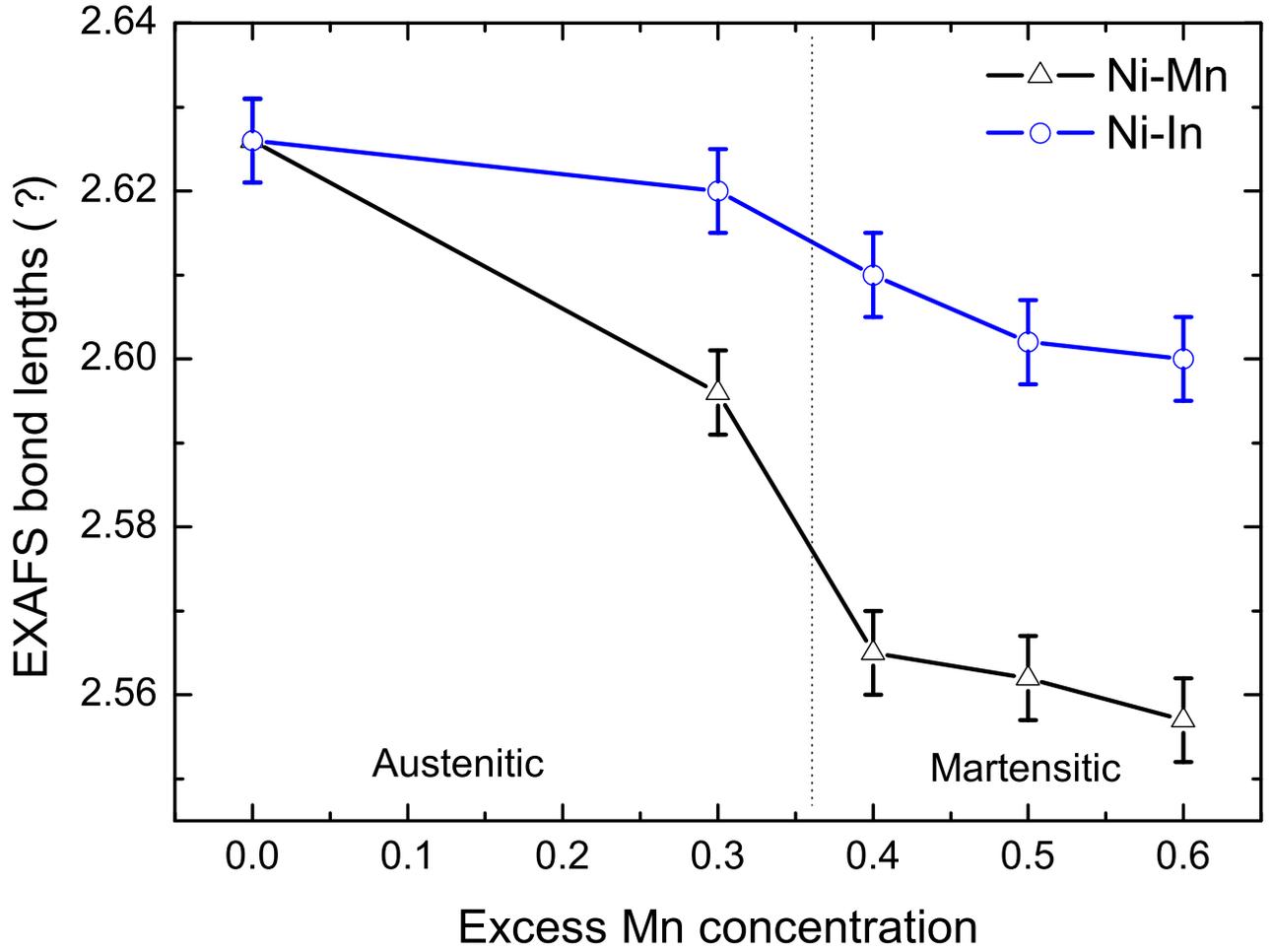}
\caption{\label{Ni-Mn_bondlength_variation} Variation of Ni-Mn  and Ni-In bond lengths obtained from EXAFS
analysis as a function of excess Mn concentration in Ni$_2$Mn$_{1+x}$In$_{1-x}$ alloys.}
\end{figure}

The variation of bond distances as a function of $x$ is plotted in Fig. \ref {Ni-Mn_bondlength_variation}. It is
clearly seen that with increasing Mn concentration the Ni-Mn or Mn-Ni bond distances decrease much more than
Ni-In bond distances and there is a sudden contraction in Ni-Mn bond length for compositions that exhibit
martensitic transformation. In all these compositions the Ni-Mn bond distance is of the order of 2.56 \AA. It may
be recalled that in Ni$_2$MnGa, Ni-Mn bond distance is about 2.54 \AA \cite{pab}. The closeness of these two
values indicate that indeed chemical pressure is responsible for an alloy to undergo martensitic transformation.
It may be recalled that In is much larger than Ga while Mn has approximately the same size. However, in these
alloys, the Ni-Mn bond distance is the average of distance between Ni and Mn atoms occupying the Y site
Ni-Mn(Y)and Ni and Mn occupying the Z-sites, Ni-Mn(Z). Since the crystal structure in the austenitic phase is
L2$_1$,  the Ni-Mn(Y) bond length is expected to be similar to Ni-In. Therefore, in these non-stoichiometric
Ni-Mn-In alloys chemical pressure is exerted by the excess Mn ion which replaces In at the Z-site in L2$_1$
Heusler structure. The Mn ion being smaller in size compared to In, it is displaced (by $\sim$ 0.06 \AA) from its
crystallographic position closer to Ni. This will result in a stronger hybridization between Ni and Mn. The band
structure calculations on these alloys suggest that density of states (DOS) near Fermi level is dominated by Ni
3d and Mn 3d bands \cite{sasi}. A stronger hybridization between Ni and Mn can result in increased DOS inducing
the material to undergo martensitic transformation. The increase in Mn concentration results in higher in the
chemical pressure and therefore higher martensitic transformation temperature.

In conclusion, the Ni and Mn K edge EXAFS show that Mn doped alloys a local structural distortion exists around
the Mn atoms even in the austenitic phase. As a result of this distortion the substituted Mn atoms are displaced
from their crystallographic positions closer to Ni giving rise to a Ni-Mn hybridization which is responsible for
inducing martensitic transformation in these Heusler alloys.

 The work at Photon Factory was performed under the Proposal No. 2009G214. Two of us (KRP and DNL) are grateful
 to Prof A. K. Raychaudhuri and DST for travel funding. Financial assistance from CSIR, New Delhi
 under the project 03(1100)/07/EMR II is acknowledged. PAB would like to acknowledge JSPS for fellowship.

\end{document}